\documentclass[useAMS,usenatbib,usegraphicx]{mn2e}
\usepackage{bm}
\usepackage{times} 

\begin{document}
\title[Far-infrared properties of dust]{Far-infrared dust properties
    in the Galaxy and the Magellanic Clouds}
\author[H. Hirashita et al.]{Hiroyuki Hirashita$^{1}$\thanks{E-mail:
    hirasita@ccs.tsukuba.ac.jp},
Yasunori Hibi$^{2}$, and Hiroshi Shibai$^{2}$
\\
$^1$ Center for Computational Sciences, University of Tsukuba,
     Tsukuba 305-8577, Japan \\
$^2$ Graduate School of Science, Nagoya University, Nagoya
     464-8602, Japan
}
\date{2007 March 19}
\pubyear{2007} \volume{000} \pagerange{1}
\twocolumn

\maketitle \label{firstpage}
\begin{abstract}
A recent data analysis of the far-infrared (FIR) map of
the Galaxy and the Magellanic Clouds has shown that there
is a tight correlation between two FIR colours: the
$60~\mu{\rm m}-100~\mu{\rm m}$ and
$100~\mu{\rm m}-140~\mu{\rm m}$ colours.
This FIR colour relation called ``main correlation'' can be
interpreted as indicative of a sequence of various
interstellar radiation fields with a common FIR optical
property of grains. In this paper, we constrain the FIR optical
properties of grains by comparing the calculated
FIR colours with the observational main correlation.
We show that neither of the ``standard'' grain species
(i.e.\ astronomical silicate and
graphite grains) reproduces the main correlation.
However, if the emissivity index at
$100~\mu{\rm m}\la\lambda\la 200~\mu$m is
changed to $\simeq 1$--1.5 (not $\simeq 2$ as the above two
species), the main correlation can be successfully
explained.
Thus, we propose that the FIR emissivity index is
$\simeq 1$--1.5 for the dust in the Galaxy and the Magellanic
Clouds at $100~\mu{\rm m}\la\lambda\la 200~\mu{\rm m}$.
We also consider the origin of the minor correlation
called ``sub-correlation'', which can be used to estimate the
Galactic star formation rate.
\end{abstract}

\section{Introduction}

Dust grains absorb stellar ultraviolet (UV)--optical light
and reprocess it into far-infrared (FIR), thereby affecting
the energetics of interstellar medium (ISM)
\citep[e.g.][]{hirashita02}. The FIR luminosity is known
to be a good indicator of star formation rate (SFR) in
galaxies \citep{kennicutt98,inoue00,iglesias04}. This can
be explained if a large part of dust grains are heated by
UV light from massive stars. The FIR spectral energy
distribution (SED) of dust grains reflects various
information on the grains themselves and on the sources of
grain heating \citep[e.g.][]{takeuchi05,dopita05}. Dust
temperature, which can be estimated from wavelength at the
peak of FIR SED, is determined by the intensity of the
interstellar radiation field (ISRF) and the optical
properties of dust
(i.e.\ how it absorbs and emits light). Since dust
grains absorb UV light efficiently, the FIR luminosity
and the dust temperature mostly reflect the UV radiation field
\citep{buat96}.

The FIR SED of the Galaxy (Milky Way) -- the ``nearest''
galaxy -- has been investigated by various authors. For
example, \citet{desert90} and \citet{dwek97} explain the
Galactic FIR SED as well as the extinction curve.
\citet{li97} provide a unified explanation for the
Galactic polarization and extinction curve by adopting
grains composed of silicate cores and organic refractory
mantles, small carbonaceous particles, and polycyclic
aromatic hydrocarbons (PAHs). Although there are some
variations for assumed species, three grain components are
usually assumed: silicate, graphite and PAHs.
Besides the grain composition and the
ISRF, the size distribution also affects the FIR SED.
The temperature of silicate and graphite
grains stays at an equilibrium temperature determined by the
radiative equilibrium if the size is roughly larger than
0.01 $\mu$m \citep{draine85}. Such large grains are called
large grains (LGs). Since the equilibrium temperature is
usually $\sim 15$--30 K in galactic environments,
the emission peak lies around wavelengths of
$\sim 100$--$200~\mu$m. If grains are smaller,
they are
transiently (stochastically) heated to a large
temperature and emit
photons with wavelengths shorter than $\sim 60~\mu$m
\citep{draine85}.
Those grains are called very small grains (VSGs).
The other component, PAH, only contributes to
wavelengths shorter than $\sim 20~\mu$m. Since we are
interested in wavelengths longer than
$\sim 60$ $\mu$m, we do not treat PAHs in this paper.

So far, a major part of works have focused on FIR
emission from high Galactic-latitude cirrus or from
some limited regions in the Galactic plane
\citep[e.g.][]{dwek97,lagache98}. Because of the
complication in the Galactic plane, which has large
inhomogeneity with a variety of ISRF
intensity and with a large range of gas (dust) density,
etc., one may expect that a unified understanding of
FIR SED in the Galactic plane is impossible. However,
\citet[][hereafter H06]{hibietal06} show that the FIR
colour-colour relation
(60--100 $\mu$m vs.\ 140--100 $\mu$m) defined by data
points with Galactic latitudes of
$|l|\leq 5^\circ$ has a clear sequence. A large fraction
($\sim 90$--95\%) of
the data points are on a colour-colour relation which they
call ``main correlation''. A minor part ($\sim 5$--10\%)
of points belong to another sequence called
``sub-correlation''
\citep[see also][]{sakon04,sakon07}. The sub-correlation tends
to be found in regions with large radio intensity,
which indicates a strong ISRF.

The ``main correlation'' of the Galactic plane is also
consistent with the FIR colours of a sample of nearby
galaxies compiled by \citet{nagata02} (see Fig.\ 8 of
H06). In particular, the FIR colour-colour relation
(60--100 $\mu$m vs.\ 140--100 $\mu$m) of the Magellanic
Clouds is located at the extension of the main
correlation of the Galactic plane. As stated by H06,
this implies that the optical properties of grains in
FIR are common
among the Galaxy and the Magellanic Clouds. H06 also
show that the FIR colours predicted
by \citet{li01} based on ``standard'' grain
optical properties \citep{draine84}
cannot reproduce the main correlation.

Indeed, some observational and experimental results
show that the FIR grain properties may be different
from the ``standard'' ones. Based on the data of the
Far-Infrared Absolute Spectrophotometer (FIRAS)
aboard the {\it Cosmic Background Explorer}
({\it COBE}), \citet{reach95} suggest that the
Galactic dust has more enhanced emissivity at
$\lambda\ga 100~\mu$m than expected before. Some
ISOPHOT observations have also reached a similar
conclusion \citep{delburgo03}, although
those observation target dense regions, where the
dust properties may be modified by grain coagulation
to form fluffy
aggregates and/or by coating of ice mantles.
Recently, \citet{dobashi05} have shown that the
extinction at
$V$ band derived from the FIR dust emission is
higher than that derived from the optical
Digitized Sky Survey data, suggesting a previous
underestimation of dust emissivity at FIR wavelengths.
An enhanced FIR emissivity is also derived for
extragalactic objects \citep{dasyra05}.
\citet{agladze96} find experimentally that
some kinds of amorphous silicate show emissivities
larger than those found in \citet{draine84}.
Moreover, the wavelength dependence suggested by
those authors is different from that presented by
\citet{draine84}. Therefore, it is worth reexamining
the FIR emissivity of dust by using
the new observational results provided by H06.

The aim of this paper is to further investigate the
origin of the main correlation both observationally
and theoretically. Since the same main correlation is
defined for the Magellanic Clouds, we also treat the
FIR emission in these galaxies. Furthermore,
we extend the analysis of H06 to
high-Galactic-latitude regions to examine if the
universality of the main correlation found in the
Galactic plane holds over the entire Galaxy. Since the
ISRF in high Galactic latitudes
is considered to be more uniform than in the
Galactic plane, the dependence on the Galactic
latitude provides us with a key to understand the
effect of inhomogeneous ISRF. A theoretical
framework for the FIR SED developed by
\citet[][hereafter DL01]{draine01} is adopted for
theoretical analysis of the main correlation.
The sub-correlation is also investigated.

This paper is organized as follows. First, in
section \ref{sec:model}, we describe the models adopted
to calculate the FIR SED. Then, in section \ref{sec:data},
we briefly review the observational results by H06,
presenting also some additional analyses. In section
\ref{sec:results}, the FIR colour-colour relation
predicted by our models
is compared with the observational main correlation.
Here, the sub-correlation is also discussed.
Section \ref{sec:sum}
is devoted to the conclusion.

\newpage

\section{FIR SED Model}\label{sec:model}

We adopt the theoretical framework of DL01 to calculate
the SED of dust emission. Since we are interested in a
wavelength range of $\lambda\sim 60$--140 $\mu$m, we
neglect PAHs, which contribute to the emission at
$\lambda\la 20~\mu$m
(\citealt{dwek97}; DL01). Our assumptions on the
relevant quantities are described below.

\subsection{Interstellar radiation field}
\label{subsec:isrf}

The ISRF in the solar neighborhood is used as the
standard. The spectrum of the ISRF in the solar
neighborhood is modeled by \citet*{mathis83}. The mean
ISRF in the solar neighborhood integrated for all the
solid angle is denoted as
$4\pi J_\lambda^\odot$. Multiplying it with $\lambda$,
\citet{mathis83} give the following numerical fitting
in units of erg cm$^{-2}$ s$^{-1}$:
\begin{eqnarray}
4\pi\lambda J_\lambda^\odot =\left\{
\begin{array}{l}
0 ~~~~~ (\lambda_{\mu{\rm m}}\leq 0.0912) \\
38.57\lambda_{\mu{\rm m}}^{4.4172} \\ ~~~~~~~
(0.0912<\lambda_{\mu{\rm m}}\leq 0.110) \\
2.045\times 10^{-2}\lambda_{\mu{\rm m}}\\ ~~~~~~~
(0.110<\lambda_{\mu{\rm m}}\leq 0.134) \\
7.115\times 10^{-4}\lambda_{\mu{\rm m}}^{-0.6678}\\
~~~~~~~
(0.134<\lambda_{\mu{\rm m}}\leq 0.246) \\
4\pi\lambda_{\mu{\rm m}}[W_1B_\lambda (T_1)+
W_2B_\lambda (T_2)+W_3B_\lambda (T_3)]\hspace{-3cm}\\
 ~~~~~~~ (0.246<\lambda_{\mu{\rm m}}) \, ,
\end{array}
\right.\label{eq:isrf}
\end{eqnarray}
where $\lambda_{\mu{\rm m}}$ is the wavelength in
units of $\mu$m, $B_\lambda (T)$ is the
Planck function,
$(T_1,\, T_2,\, T_3)=(7500~{\rm K},\, 4000~{\rm K},\,
3000~{\rm K})$, and
$(W_1,\, W_2,\, W_3)=(1.0\times 10^{-14},\,
1.0\times 10^{-13},\, 4.0\times 10^{-13})$.
The ISRF intensity $J_\lambda$ is scaled proportionally
to $J_\lambda^\odot$ as follows:
\begin{eqnarray}
J_\lambda =\chi J_\lambda^\odot \, ,\label{eq:j_l}
\end{eqnarray}
i.e.\ the spectral shape of the ISRF is fixed and
scaled with $\chi$ ($\chi =1$ corresponds to the
ISRF intensity in the solar neighborhood).

\subsection{Optical constants}

We assume a grain to be spherical with a radius of $a$.
The absorption cross section of the grain is expressed
as $\pi a^2Q_{\rm abs}(\lambda )$, where
$Q_{\rm abs}(\lambda )$ is called absorption
efficiency. The optical constants of silicate and
graphite grains are taken from \cite{draine84}.
For the UV regime, we adopt ``smoothed astronomical
silicate'' from \citet{weingartner01}
for silicate grains.
For graphite grains, two-thirds of the particles have
an dielectric function $\epsilon =\epsilon_\bot$
and one-third $\epsilon =\epsilon_\parallel$
\citep[see][]{draine84}.

\subsection{Grain size distribution}\label{subsec:size}

The number density of grains with sizes between $a$ and
$a+{\rm d}a$ is denoted as $n_i(a)\,{\rm d}a$, where
the subscript $i$ denotes a grain species
($i=\mbox{sil or gra}$ stands for silicate and graphite,
respectively).
We assume a power-law form for $n_i(a)$:
\begin{eqnarray}
n_i(a)={\cal C}_ia^{-K}~~~(a_{\rm min}\leq a\leq
a_{\rm max})\, ,
\label{eq:size_dist}
\end{eqnarray}
where $a_{\rm min}$ and $a_{\rm max}$ are the upper and
lower cutoffs of grain size, respectively, and
${\cal C}_i$ is the normalizing constant.
\citet*{mathis77} find that the
extinction curve observed in the solar neighborhood
is reproduced by a composite model of graphite and
silicate grains with $K=3.5$.
We assume $a=3.5$ \AA\ and $a_{\rm max}=0.25~\mu$m
for both
graphites and silicates \citep{mathis77,li01}, although
\citet{li01} adopt more elaborate functional form.
We have confirmed that the difference does not
affect the following discussion on the FIR colours.

The mixing ratio between silicate and graphite is
uncertain and different even among models for the
Galactic dust \citep*{dwek97,takagi03,li01}.
According to \citet{pei92}, the ratio between those
two species is different among the Galaxy, the Large
Magellanic Cloud (LMC) and the Small Magellanic Cloud
(SMC). In the SMC, \citet{pei92} suggest a
silicate-dominated dust composition, while
\citet{welty01} argue against
such a composition from a study of the depletion
pattern \citep[but see][]{li06}. Considering such
large uncertainty in the grain composition, we present
the FIR SED of each grain
species separately to avoid the uncertainty in the
mixing ratio between the species.
The effects of mixing two species are
briefly mentioned in section \ref{subsec:mixing}.

As long as we treat the FIR colours of a single species
(section \ref{subsec:colour}), the constant
${\cal C}_i$ cancels out. Thus, it is not necessary to
determine the absolute abundance of graphite and
silicate. The relative ratio of ${\cal C}_i$ between
species is
important only when we consider a mixture of
various species (section \ref{subsec:mixing}).
In this case, ${\cal C}_i$ is determined by
\begin{eqnarray}
\rho_i=
\int_{a_{\rm min}}^{a_{\rm max}}\frac{4\pi}{3}a^3
s_i{\cal C}_ia^{-K}\,{\rm d}a\, ,
\end{eqnarray}
where $\rho_i$ is the mass density of species $i$
in the ISM, and $s_i$ is the material density
of species $i$. We assume $s_{\rm sil}=3.3$ g cm$^{-3}$
and $s_{\rm gra}=2.2$ g cm$^{-3}$ \citep*{jones96}.
In fact, we require only the ratio
${\cal C}_{\rm sil}/{\cal C}_{\rm gra}$
(or $\rho_{\rm sil}/\rho_{\rm gra}$) to calculate
the FIR colors in a mixture of the two species.

\subsection{Heat capacity}\label{subsec:cap}

We adopt multi-dimensional Debye models for the heat
capacity of grains. The heat capacity of a graphite
grain per unit volume is (DL01; \citealt{takeuchi03})
\begin{eqnarray}
C_{\rm gra}=(N_{\rm C}-2)k_{\rm B}\left[
f_2'\left(\frac{T}{863~{\rm K}}\right)+2f_2'\left(
\frac{T}{2504~{\rm K}}\right)\right]\, ,
\end{eqnarray}
\begin{eqnarray}
f_n(x)\equiv n\int_0^1\frac{y^n{\rm d}y}{\exp (y/x)-1}\, ,
~~~f_n'(x)\equiv\frac{\rm d}{{\rm d}x}f_n(x)\, ,
\end{eqnarray}
where $N_{\rm C}=1.14\times 10^{23}$ cm$^{-3}$
\citep{takeuchi03} is the number of carbon atoms
contained in the unit volume of the grain, $k_{\rm B}$ is
the Boltzmann constant, and $T$ is the grain temperature.
The heat capacity of a silicate grain per unit volume is
(DL01)
\begin{eqnarray}
C_{\rm sil}=(N_{\rm a}-2)k_{\rm B}\left[2f_2'\left(
\frac{T}{500~{\rm K}}\right)+f_3'\left(
\frac{T}{1500~{\rm K}}\right)\right]\, ,
\end{eqnarray}
where $N_{\rm a}=8.5\times 10^{22}$ cm$^{-3}$ is the
number of atoms contained in the unit volume of the grain
\citep{takeuchi03}.

\subsection{Calculation of FIR colours}\label{subsec:colour}

The total FIR intensity (per frequency per hydrogen nucleus
per solid angle) of dust emission at a wavelength $\lambda$
can be estimated as
\begin{eqnarray}
I_\nu^i(\lambda )=\int_{a_{\rm min}}^{a_{\rm max}}{\rm d}
a\frac{1}{n_{\rm H}}n_i(a)\pi a^2
Q_{\rm abs}(\lambda )\int_0^\infty{\rm d}T\, B_\nu (T)
\frac{{\rm d}P_i}{{\rm d}T},\hspace{-1cm}\nonumber\\
\end{eqnarray}
where $i$ specifies the grain species and
${\rm d}P_i/{\rm d}T$ is the temperature distribution
function of the grains. We assume that the FIR radiation
is optically thin. Thus, the observed colour directly
reflects the ratio of the above intensity. Here the
colour at wavelengths
of $\lambda_1$ and $\lambda_2$ is defined as
$I_\nu (\lambda_1)/I_\nu (\lambda_2)$, which is called
$\lambda_1-\lambda_2$ colour.

The distribution function ${\rm d}P_i/{\rm d}T$ is
calculated by the formalism of DL01 (see their section 4)
based on the quantities in sections
\ref{subsec:isrf}--\ref{subsec:cap}. The enthalpy of a
grain is divided into discrete bins,
and the transition probability between the bins is
treated by taking into account the heating by
interstellar radiation and the cooling by emission
\citep[see also][]{guhathakurta89}.
 
\section{Data}\label{sec:data}

\subsection{Outline of the analysis by H06}

The data analyzed by H06 are adopted in this paper.
H06 used the Zodi-Subtracted Mission Average (ZSMA)
taken by the Diffuse Infrared Background Experiment
(DIRBE) of the {\it COBE}. They analyzed the data at three
FIR wavelengths: $\lambda =60~\mu$m, 100 $\mu$m, and
140 $\mu$m. Then, they examined the relation between
two colours,
$I_\nu (60~\mu{\rm m})/I_\nu (100~\mu{\rm m})$
and $I_\nu (140~\mu{\rm m})/I_\nu (100~\mu{\rm m})$.
Following H06, we adopt the intensity after colour
correction for the 100-$\mu$m and 140-$\mu$m bands.
For the Galactic plane, they treated regions with
Galactic latitudes of $|b|<5^\circ$.
The FIR data of the Magellanic Clouds are also
analyzed. In order to avoid the uncertainty
in the subtraction of the zodiacal component,
they only select the region with 
$I_\nu (60~\mu{\rm m})>3$ MJy sr$^{-1}$.
The error of $I_\nu$ is less than 10\%.

\begin{figure*}
\begin{center}
\includegraphics[width=8cm]{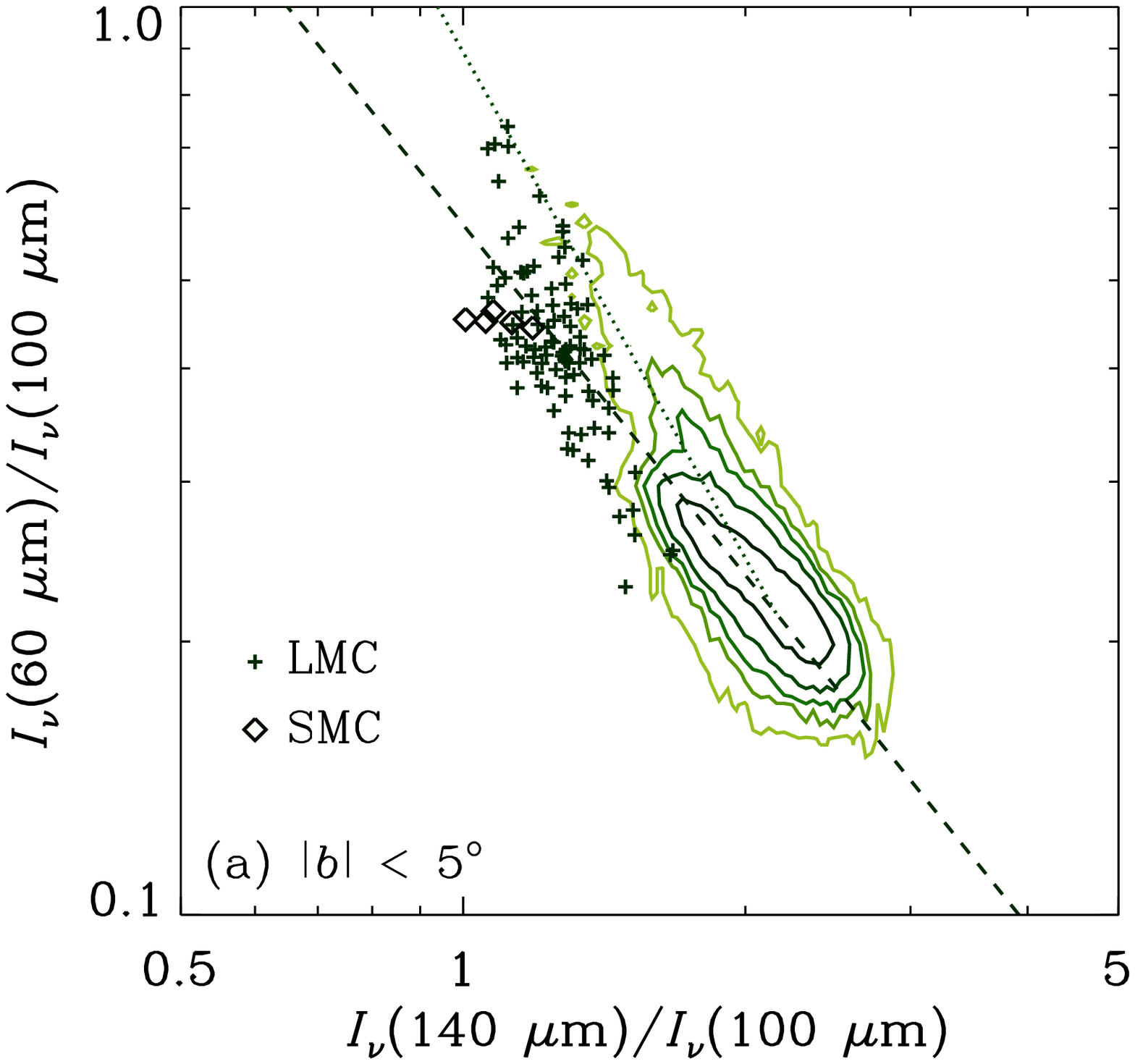}
\includegraphics[width=8cm]{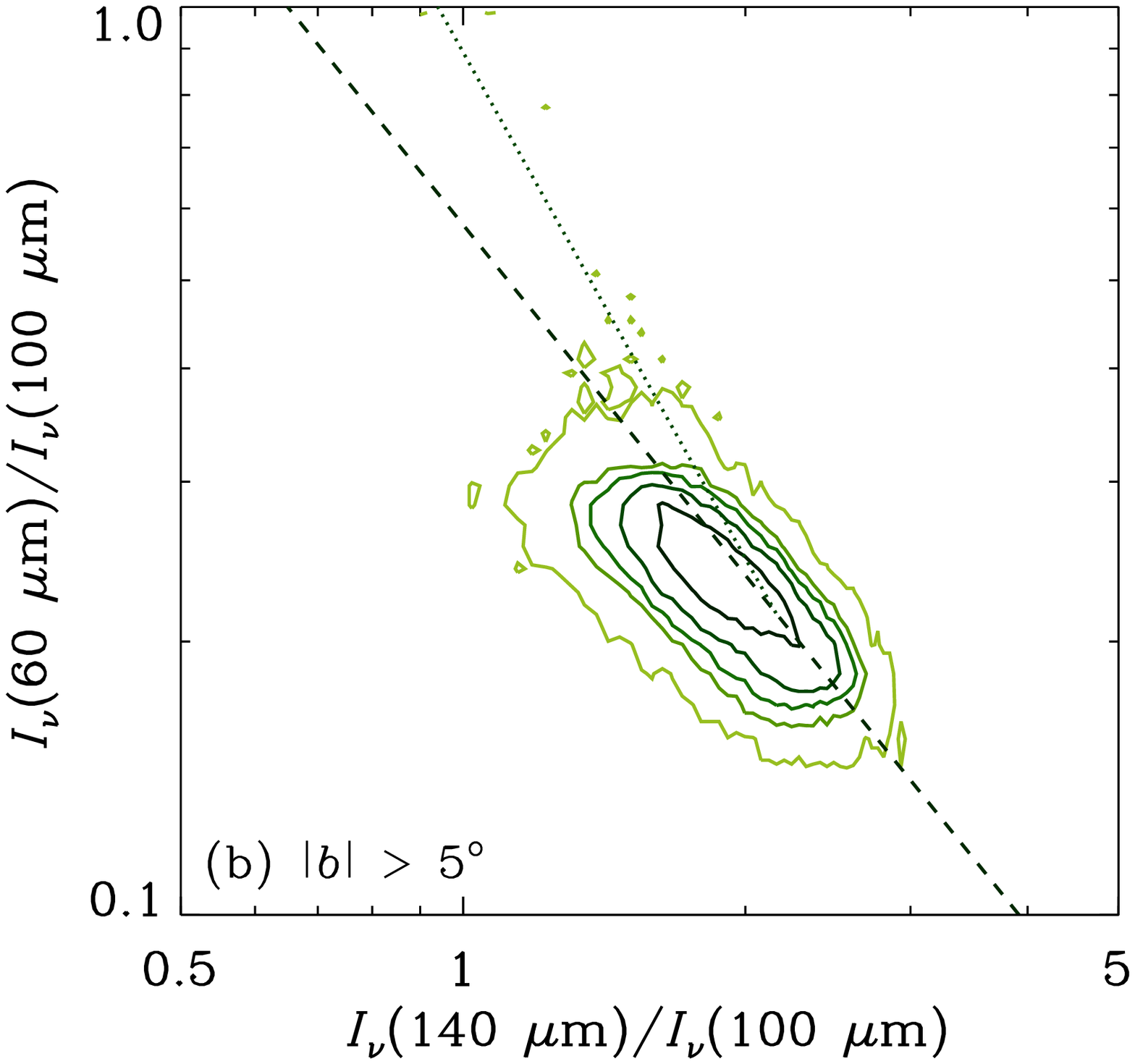}
\end{center}
\caption{Far-infrared colour-colour
($60~\mu{\rm m}-100~\mu{\rm m}$ colour vs.\
$140~\mu{\rm m}-100~\mu{\rm m}$ colour) diagram derived from
the COBE/DIRBE ZSMA map data. (a) The data of the Galactic
plane ($|b|<5^\circ$), the LMC, and the SMC are shown.
The contours show
the distribution of the Galactic plane data.
The contours show the levels where 50\%, 80\%, 90\%, 95\%
and 99\% of the data are contained.
The LMC and SMC data are plotted with the filled circles
and with the open diamonds, respectively.
The dashed line and the dotted lines
indicate the fittings for the main correlation and
the sub-correlation, respectively. (b) Distribution of
the data in high Galactic latutudes ($|b|>5^\circ$).
The levels of the contours and the dashed and dotted lines
are the same as those in Panel (a).
\label{fig:hibi_fig2}}
\end{figure*}

In Figure \ref{fig:hibi_fig2}a, we show the distribution
of the Galactic-plane data in the colour-colour diagram.
H06 found strong correlations between the two colours.
More than 90\% of the data lie on a strong
correlation called main correlation, which can be fitted
as (H06)
\begin{eqnarray}
\frac{I_\nu (140~\mu{\rm m})}{I_\nu (100~\mu{\rm m})}
=0.65\left(
\frac{I_\nu (60~\mu{\rm m})}{I_\nu (100~\mu{\rm m})}
\right)^{-0.78}\, .
\end{eqnarray}
The main correlation also explains the FIR colours of
the LMC and the SMC (H06b). There is another correlation
sequence, called sub-correlation in H06
\citep[see also][]{hibi06}:
\begin{eqnarray}
\frac{I_\nu (140~\mu{\rm m})}{I_\nu (100~\mu{\rm m})}
=0.93\left(
\frac{I_\nu (60~\mu{\rm m})}{I_\nu (100~\mu{\rm m})}
\right)^{-0.56}\, .
\end{eqnarray}

H06 also examined the dependence on the Galactic longitude
$l$, showing that the FIR colours tend to shift downwards
along the main correlation
as $l$ increases from 0$^\circ$ to 180$^\circ$.
Since the ISRF in the region toward the Galactic
centre tends to be
higher than that in the opposite direction
\citep{mathis83},
the colour shift along the main correlation
can be caused by the difference in ISRF; that is,
the FIR colours of dust grains irradiated by a high
ISRF tends to be situated in an upper part of
the main correlation on the colour-colour
diagram.

We extend our analysis to high Galactic
latitudes ($|b|>5^\circ$). The procedure
of the data analysis and the selection criterion
for the intensity (i.e.\
$I_\nu (60~\mu{\rm m})>3$ MJy sr$^{-1}$) are the
same as H06. Most of the data are taken from
$|b|\la 40^\circ$, since the criterion for the
intensity is hard to be satisfied in higher
Galactic latitudes.
In Figure \ref{fig:hibi_fig2}b,
we show the results. Although the data distribution
is shifted slightly downward compared with the
Galactic plane data, we clearly see the main
correlation. Moreover, the sub-correlation is
not clear in the high Galactic latitudes. This
supports the idea of H06 that the sub-correlation
is produced by the contamination of high-ISRF
regions, which tend to reside in the Galactic
plane.

\subsection{The main correlation}

The data presented in Figure \ref{fig:hibi_fig2}
indicate that the main correlation is robust against
the change of Galactic latitudes ($b$). H06 also show
that the
FIR colours shifts along the main correlation by the
variation of Galactic longitudes ($l$). These facts
mean that the main correlation is independent of
the complexity of the Galactic
structure. Since the inhomogeneity of
ISRF should be smaller in high Galactic latitudes than
in the Galactic plane, the robustness of the main
correlation over a variety of $b$ suggests that a
detailed modeling of
galactic structures and of inhomogeneous ISRF are not
necessary to understand the main correlation.

As mentioned in the previous subsection, H06 argued
that the main correlation is produced by a sequence of
varying ISRF: the FIR colours tend to be located in an
upper part of the main correlation for a high ISRF.
However, this was only suggested from the longitude
dependence but was not
theoretically demonstrated. Thus, in
section \ref{sec:results}, we
investigate how the FIR colour-colour relation
changes for various ISRF intensities. By examining if
the observed colour-colour relation is explained or not,
we will finally be able to
test what kind of grain species is consistent with
the observational FIR colour-colour relation
(section \ref{subsec:fir_opt}).

It is worth noting that the main correlation can also
explain the FIR colours in the Magellanic Clouds.
Since the LMC has a face-on geometry, the inhomogeneity
in the ISRF in a line of sight may be small. The fact
that the FIR colour-colour
relation of the LMC shifts upward relative to that of
the Galaxy suggests that the location in the
main correlation reflects the strength of ISRF, since
the LMC is generally believed to have a higher ISRF than
the Galaxy \citep[e.g.][]{fukui01,tumlinson02}. It is
surprising that the FIR colours of
the LMC and the SMC are located on the main correlation
defined by the Galactic dust grains, and it is
interesting to give a unified understanding of the
FIR colours for those three galaxies.
We also tackle this problem in
sections \ref{sec:results} and \ref{sec:discussion}.

\subsection{The sub-correlation}

H06 show that the sub-correlation can be reproduced by
summing two colours in the main correlation.
Thus, they argue that the main correlation is fundamental.
We also follow their argument and concentrate on explaining
the main correlation. The sub-correlation is
investigated in section \ref{subsec:mixture}.

\section{Results}\label{sec:results}

\subsection{Standard grain optical constants}
\label{subsec:std}

{}From various observational evidence such as extinction
curves and interstellar depletion patterns,
dust is believed to be mainly composed of two species:
silicate and carbonaceous dust
\citep{mathis90,li97,jones00,okada06}. For those two
species, \citet{draine84} proposed astronomical silicate
and graphite, which reproduce the observational
extinction curve and FIR emission spectrum
\citep[see also][]{li01}.

\begin{figure}
\begin{center}
\includegraphics[width=8cm]{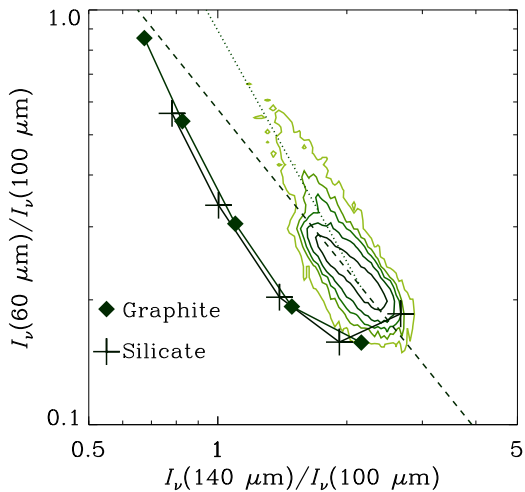}
\end{center}
\caption{FIR colour-colour
($60~\mu{\rm m}-100~\mu{\rm m}$ vs.\
$140~\mu{\rm m}-100~\mu{\rm m}$) relation
for silicate and graphite grains (large crosses and
large squares,respectively). The different points
correspond to
$\chi =0.3$, 1, 3, 10, and 30 from low to high
$140~\mu{\rm m}-100~\mu{\rm m}$ colours.
The same contour as shown in
Figure \ref{fig:hibi_fig2}a as well as the
main and sub-correlations (dashed and dotted lines,
respectively) are also presented.
\label{fig:clr_std}}
\end{figure}

Based on the models described in section \ref{sec:model},
we examine how the FIR colours change by the variation of
ISRF ($\chi$). Following H06, we focus on the relation
between the 60 $\mu$m--100 $\mu$m colour
[$I_\nu (60~\mu{\rm m})/I_\nu (100~\mu{\rm m})$] and
the 140 $\mu$m--100 $\mu$m colour
[$I_\nu (140~\mu{\rm m})/I_\nu (100~\mu{\rm m})$].
In Figures \ref{fig:clr_std}, we plot the colour-colour
relations of graphite and astronomical silicate for
$\chi =0.3$, 1, 3, 10, and 30. The discrepancy between
the observational main correlation
and the theoretical prediction is significant.
Some mechanism that shifts
the colour-colour relation upward and/or rightward should be
included to
explain the observational colour correlations.

It should be noticed that both silicate and graphite
grains trace similar tracks on the FIR colour-colour
relation although the UV absorption coefficients and
heat capacities are different between those two species.
This strongly suggests that the track on the FIR
colour-colour diagram is determined mainly by the FIR
absorption coefficients of grains. Indeed, both have
dependence roughly described as
$Q_{\rm abs}\propto\lambda^{-2}$. Thus, we suspect that
the FIR colour-colour relation is not sensitive to
various factors other than the wavelength dependence of
the FIR absorption coefficient.

\begin{figure*}
\begin{center}
\includegraphics[width=7.5cm]{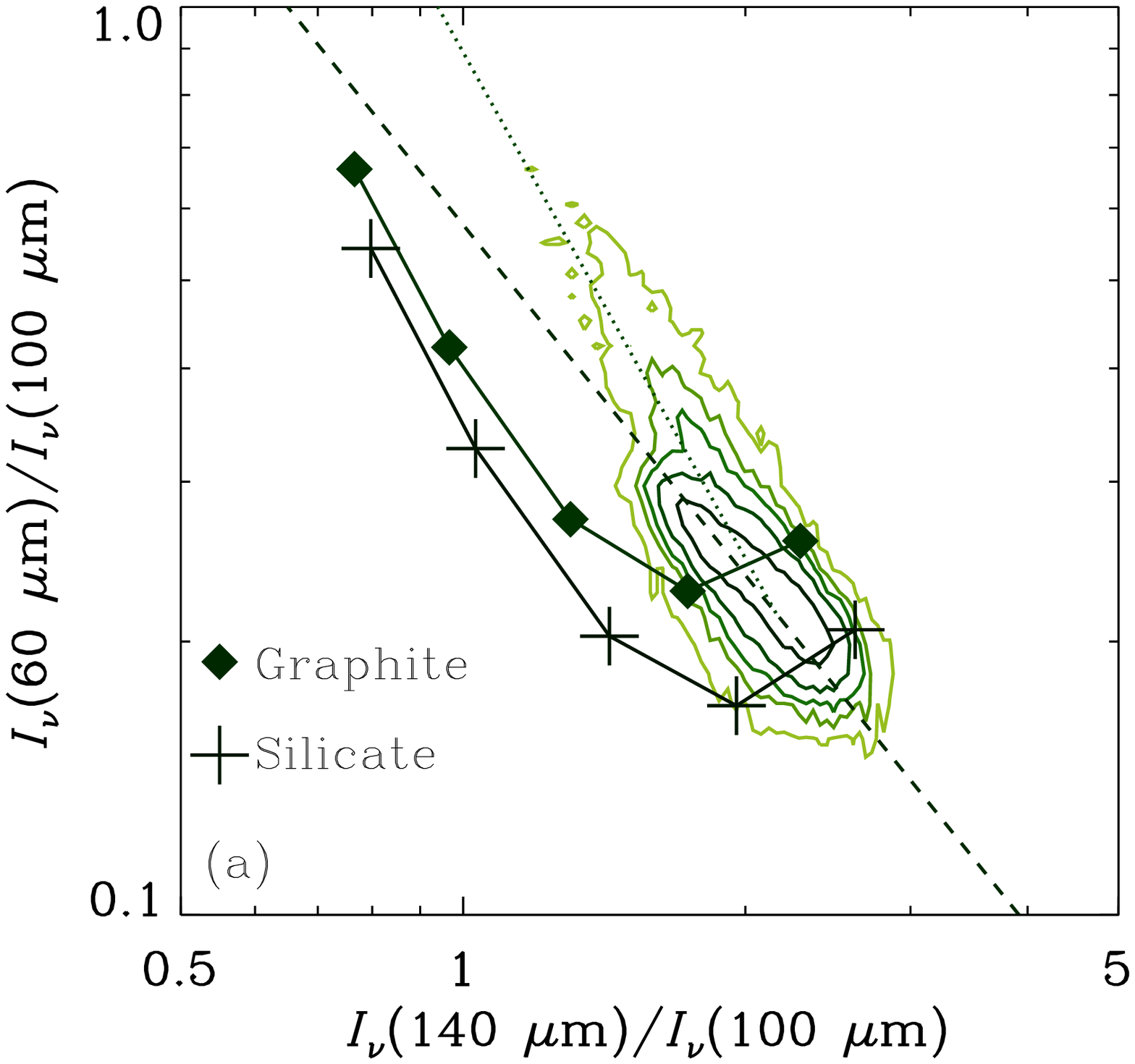}
\includegraphics[width=7.5cm]{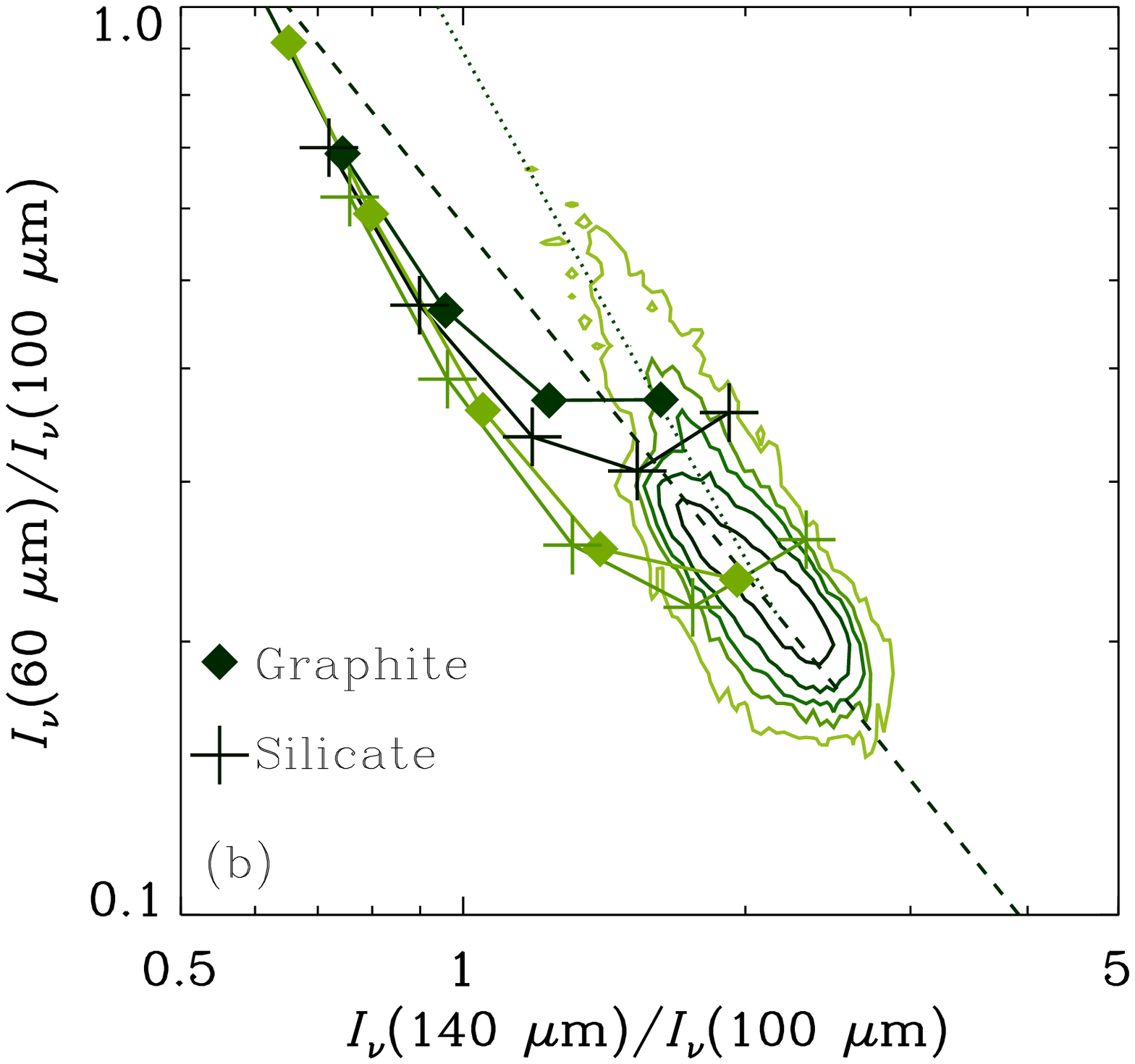}
\end{center}
\caption{Same as Figure \ref{fig:clr_std} but for
(a) the hard ISRF and (b) the size distribution
of grains with $K=3.75$ and 4.0 (lower and upper
sequences, respectively).
\label{fig:clr_enhanced60}}
\end{figure*}

The colour-colour relation shifts upwards if the
contribution from VSGs relative to that of LGs
increases, since the emission from stochastically
heated VSGs contributes significant to the emission
at $\lambda =60~\mu$m. Indeed, grains with
$a\sim 50$ \AA\ contribute to the 60 $\mu$m intensity
\citep{draine07}. A change of the ISRF spectrum
and/or of the grain size distribution can change the
relative contribution from the stochastic VSG
emission. First we vary the ISRF spectrum. A hard
spectrum could relatively enhance the emission from VSGs,
since the cross section of a VSG is sensitive to the change
of wavelength in the UV regime \citep[e.g.][]{draine84}.
We calculate the FIR colours by assuming a ``hard''
spectrum without the optical--NIR bump; i.e.\
we adopt
$7.115\times 10^{-4}\lambda_{\rm \mu m}^{-0.6678}$
for $\lambda_{\rm \mu m}>0.246$ instead of the
expression in equation (\ref{eq:isrf}).
The FIR colour-colour relation calculated with this
hard spectrum is shown in Figure \ref{fig:clr_enhanced60}a.
The colour-colour relation shifts upward and approaches
the main correlation compared with the relation in
Figure \ref{fig:clr_std}. This is due to
relative enhancement of the stochastic heating of VSGs,
which contributes to the 60 $\mu$m emission. However, the
change of the FIR colours for various $\chi$ does not
follow the main correlation.
Moreover, such a hard spectrum as
assumed here is unrealistic for high Galactic latitude,
where the main correlation is also seen. Thus,
we conclude that the change of the
UV spectrum cannot explain the observational FIR
colours.

The enhancement of the stochastic heating occurs also if
the relative number of VSGs increases. Thus, we expect
that the colour-colour relation shifts upward if we assume
a grain size distribution with an enhanced number of
VSGs. The number of VSGs may be increased by shattering
\citep{jones96}. The size distribution shown by
\citet{jones96} can be roughly fitted by
$n(a)\propto a^{-4.0}$ (i.e.\ $K=4.0$ in
equation \ref{eq:size_dist}). We also examine $K=3.75$
as an intermediate case. In both cases, we adopt the
same values of $a_{\rm min}$ and $a_{\rm max}$ as
assumed in section \ref{subsec:size}. In
Figure \ref{fig:clr_enhanced60}b, we show the
colour-colour relation for $K=3.75$ and 4.0. We observe
that the 60 $\mu$m--100 $\mu$m colour is sensitive to the
grain size distribution. However, the overall trend of
the main correlation is not reproduced neither by
$K=3.75$ nor by $K=4.0$.
The 60 $\mu$m--100 $\mu$m colour is clearly out of the
observed range for
$K=4.0$. Although $K=3.75$ explains the most concentrated
part of the
data with $\chi\sim 0.3$--1, the trend along the
main correlation is not reproduced.

The upper and lower cutoffs of grain size
($a_{\rm max}$ and $a_{\rm min}$) can also be changed.
However, the 140 $\mu$m--100 $\mu$m colour
is not
sensitive to the change of $a_{\rm min}$ because it
is determined by the equilibrium temperature of
large grains. It is not largely affected by the
change of $a_{\rm max}$ as long as
$a_{\rm max}\ga 0.1~\mu$m.
The 60 $\mu$m--100 $\mu$m
colour monotonically increases if the fraction
of VSGs increases. According to \citet{draine07},
grains with $a\sim 50$ \AA\ contribute to the
60 $\mu$m flux, while a large fraction of the
100 $\mu$m flux comes from grains with
$a\ga 100$ \AA.
Thus, as $a_{\rm min}$ decreases
and/or $a_{\rm max}$ decreases,
the 60 $\mu$m--100 $\mu$m colour monotonically
increases. However, this colour is not sensitive
to the grain size distribution at $a\ll 50$ \AA,
and the abundance of such very small particles
is constrained better in mid-infrared
\citep[e.g.][]{sakon07}.
We have confirmed after calculations
with our models that resulting colour-colour relations
predicted by decreased $a_{\rm min}$ and/or
$a_{\rm max}$ are similar to those in
Figure \ref{fig:clr_enhanced60}
and that the observed main correlation can never be
explained by changes of $a_{\rm min}$ and
$a_{\rm max}$.

In summary, it is difficult to explain the overall
main correlation by simply enhancing the stochastic
heating component, which contributes to the 60 $\mu$m
intensity. Heat capacities and grain optical constants
are also concerned with the FIR SED. However, as shown
in Appendix \ref{app:mat}, difference in the heat
capacity does not change the FIR colour-colour relation.
There, we also show that different materials with
different UV--near-infrared absorption coefficients
does not explain the observational FIR colour-colour
relation. We note that all the materials examined in
Appendix \ref{app:mat} have the FIR spectral index
(which is defined as
$\beta\equiv -{\rm d}\ln Q_{\rm abs}(\lambda )/{\rm d}\ln\lambda$,
i.e.\ $Q_{\rm abs}$ is approximated to be proportional
to $\lambda^{-\beta}$) nearly equal to 2.
We are left with changing the FIR optical properties
of grains. 
The following subsection is devoted to examining the
possibility that the FIR spectral index is different
from 2.

\subsection{Dependence on the FIR emissivity index}
\label{subsec:emissivity}

A spectral index $\beta =2$ is the most frequently
used value, which is supported by some experiments and
observations \citep[e.g.][]{hildebrand83,draine84}.
On the other hand, an analysis of the FIRAS data
by \citet{wright91} indicates that the Galactic
dust has $\beta\simeq 1.65$.
Some other observational and experimental results
also indicate
that the FIR emissivity index may be smaller than
2 \citep{reach95,aguirre03}.\footnote{H06
adopt a colour correction with a spectral index of 2
($\beta =2$), but even if the colour is corrected
by assuming $\beta =0$, the difference in the colours
is less than 10\%. Thus, the prediction with
$\beta =1$ can be compared with the results of H06
within an uncertainty of $<10$\%.}
The range of $\beta$ is roughly from
$\sim 0.9$ to 1.6 as suggested for the FIR colours of
the Galaxy and the Magellanic Clouds
\citep{reach95,aguirre03}.
Some species indeed show such a low emissivity index
\citep{agladze96}: $\beta\simeq 1.2$ for
${\rm MgO}\cdot 2{\rm SiO}_2$ (amorphous). Since there
is no available information on the optical constants
of ${\rm MgO}\cdot 2{\rm SiO}_2$ in the wavelengths
shorter than FIR, we cannot include this species
consistently into our models.
Fortunately, the resulting FIR colours are not sensitive
to the detailed wavelength dependence of the UV absorption
coefficient as shown in Appendix \ref{app:mat}.
Thus, we adopt the emissivity of astronomical silicate
or graphite at $\lambda\leq 100~\mu$m and change the
spectral index at $\lambda >100~\mu$m. That is,
we choose the following absorption efficiency for
$\lambda >100~\mu{\rm m}$:
\begin{eqnarray}
Q_{\rm abs}(\lambda ,\,\beta)=Q_{\rm abs}
(\lambda =100~\mu{\rm m})\left(
\frac{\lambda}{100~\mu{\rm m}}\right)^{-\beta}\, ,
\label{eq:beta}
\end{eqnarray}
where we adopt the value
$Q_{\rm abs}(\lambda =100~\mu{\rm m})$ from
\citet{draine84} for astronomical silicate and graphite.

\begin{figure*}
\begin{center}
\includegraphics[width=7.5cm]{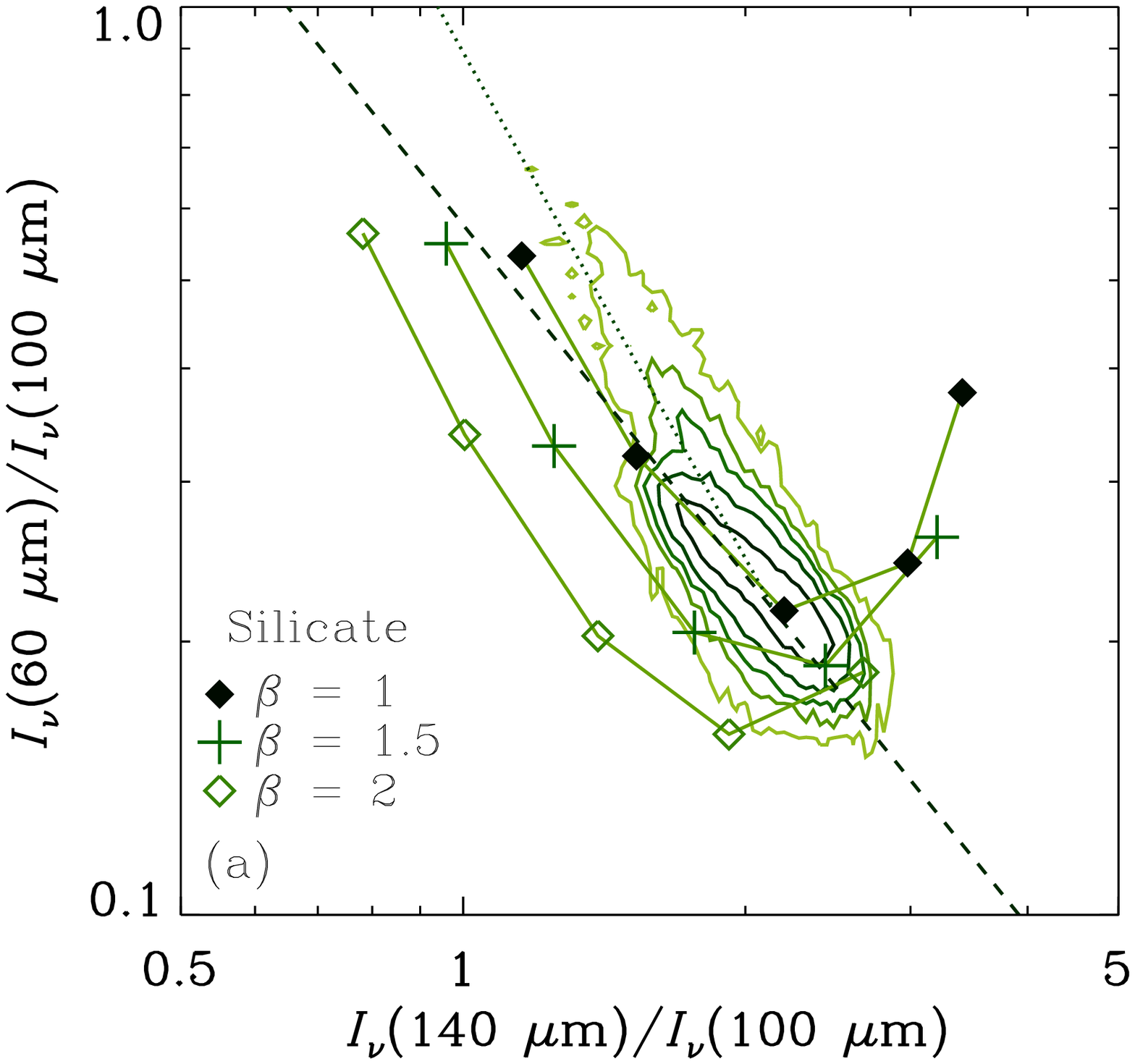}
\includegraphics[width=7.5cm]{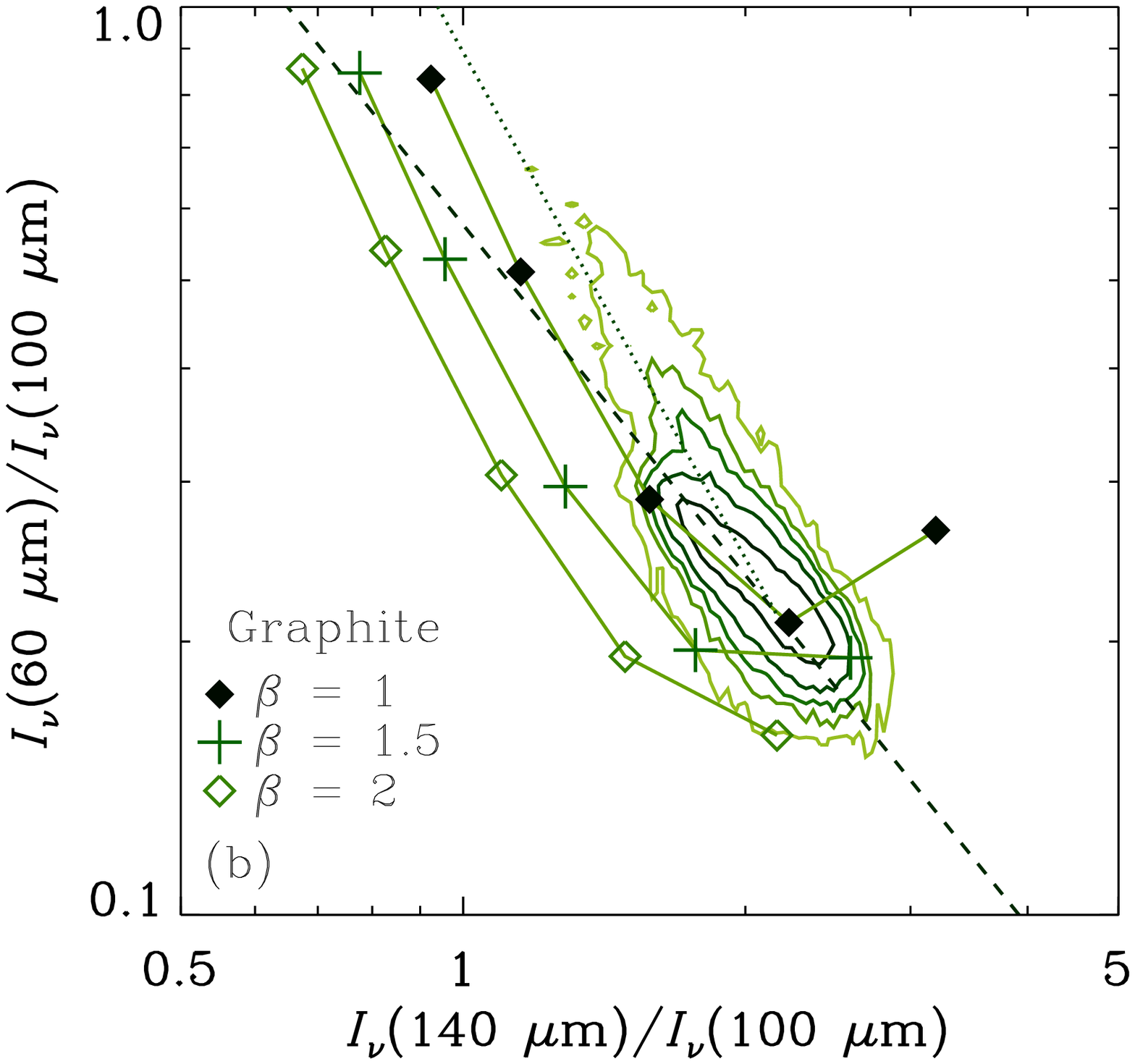}
\includegraphics[width=7.5cm]{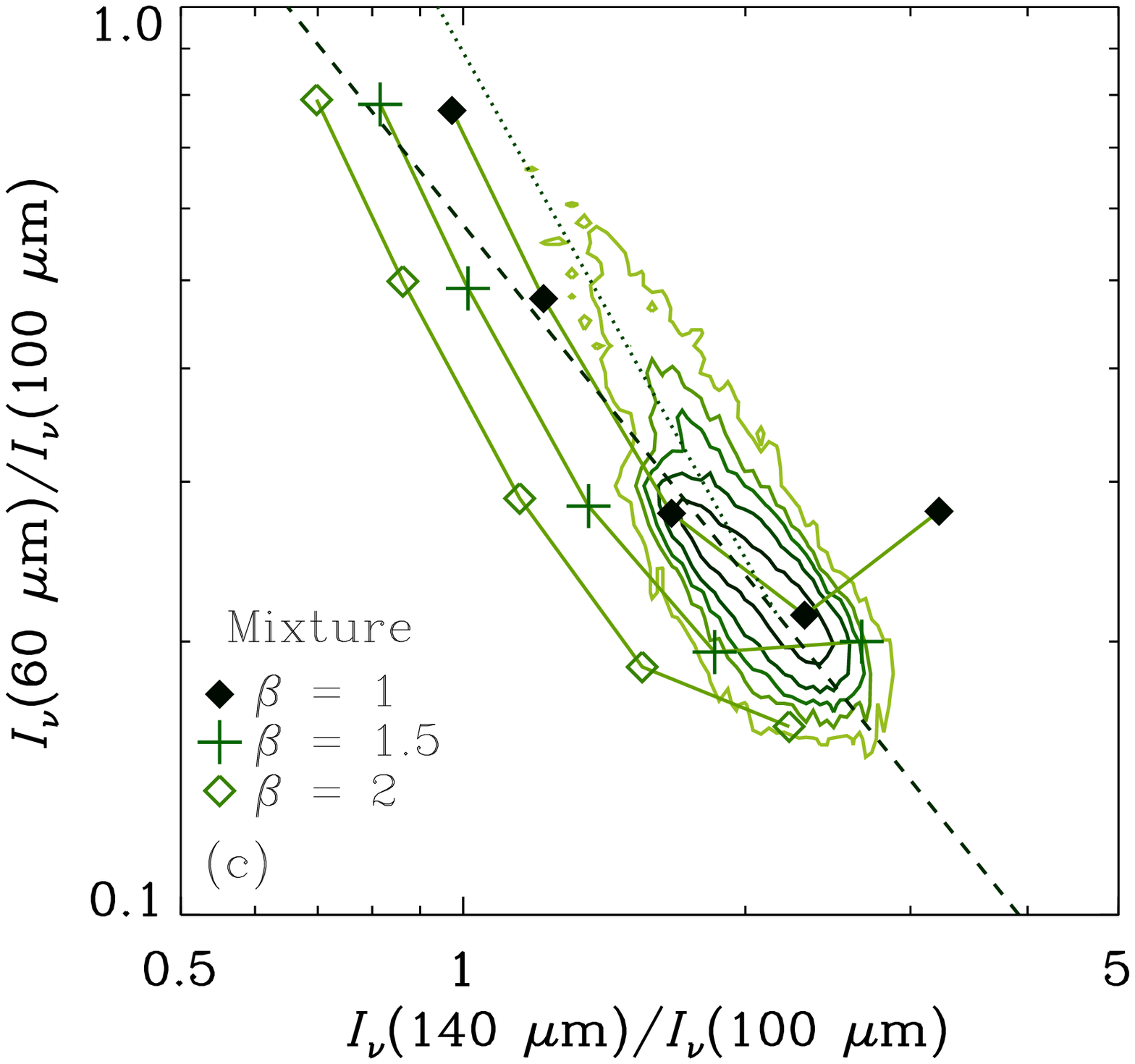}
\includegraphics[width=7.5cm]{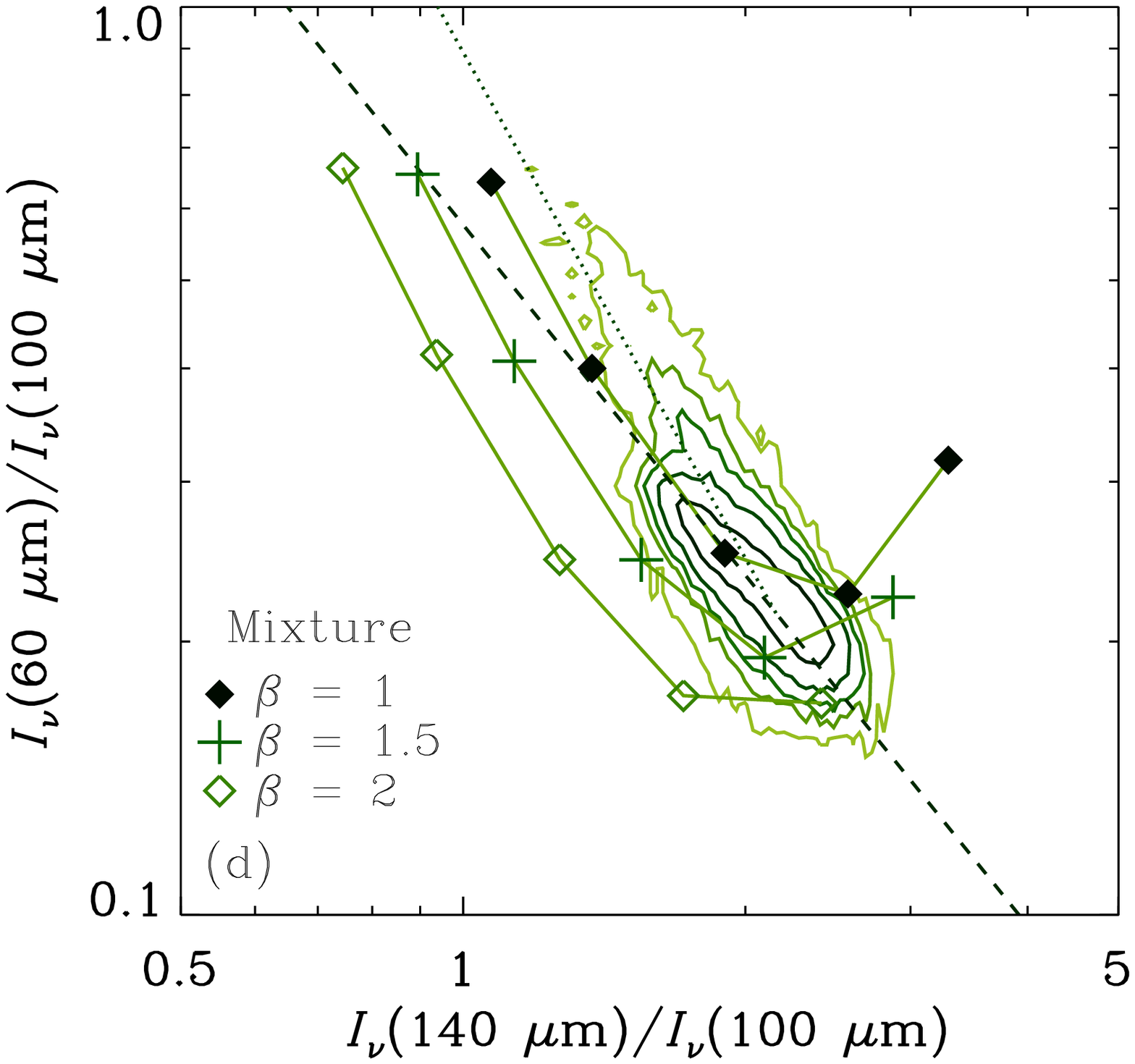}
\end{center}
\caption{Same as Figure \ref{fig:clr_std} but for
various FIR emissivity indexes $\beta$ defined at
$\lambda\geq 100~\mu$m. The filled squares, crosses, and
open squares show the FIR colour-colour relation for
$\beta =1$, 1.5, and
2, respectively. In Panels (a) and (b), we assume
the silicate and graphite absorption efficiencies
at $\lambda <100~\mu$m, respectively. In Panel (c)
and (d),
the colours-colour relation of the graphite-silicate
mixture is shown, where the mass ratios between
silicate and graphite are 1.11 : 1 and 7.25 : 1,
respectively.
\label{fig:clr_beta}}
\end{figure*}

In Figure \ref{fig:clr_beta}, we show the colour-colour
relation for various FIR spectral indexes. Here we
call the species according to the adopted optical
properties at $\lambda <100~\mu{\rm m}$ although
we modify artificially $Q_{\rm abs}$ at
$\lambda >100~\mu{\rm m}$. We observe from
Figure \ref{fig:clr_beta} that $\beta =1$ is
roughly consistent with the main correlation
if the ISRF intensity is between $\sim 1$ and
$\sim 10$ ($\beta =1.5$ may also be permitted considering
the uncertainty in the data). It is worth noting that the
colour-colour
relation is just along the main correlation.
Moreover, the region of the largest concentration of
the data (see the contours in Fig.\ \ref{fig:clr_beta})
is reproduced
with a standard ISRF $\chi\sim 1$. Thus, we conclude
that a FIR emissivity index of $\sim 1$--1.5 is
successful in reproducing the main correlation.

\begin{figure}
\begin{center}
\includegraphics[width=8cm]{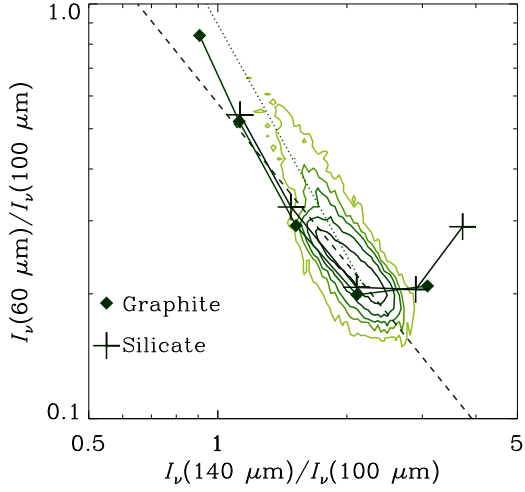}
\end{center}
\caption{Same as Figure \ref{fig:clr_std} but for
the FIR emissivity shown by equation \ref{eq:reach}.
\label{fig:clr_reach}}
\end{figure}

For the absorption efficiency at $\lambda >100~\mu$m,
we also examine a hybrid functional form proposed by
\citet{reach95}:
\begin{eqnarray}
Q_{\rm abs}(\lambda )=
\frac{(\lambda /\lambda_0)^{-2}}
{[1+(\lambda_1/\lambda)^6]^{1/6}}\, ,\label{eq:reach}
\end{eqnarray}
which behaves like equation (\ref{eq:beta})
with $\beta =1$ for
$\lambda\ll\lambda_1$ and $\beta =2$ at
$\lambda\gg\lambda_1$ \citep[see also][]{bianchi99}.
We assume that $\lambda_1=200~\mu$m, following
\citet{reach95}. For $\lambda\leq 100~\mu$m,
we adopt the same absorption efficiency of astronomical
silicate and graphite as used in
section \ref{subsec:std}. Thus, 
we set the value of $\lambda_0$ so that the continuity
of $Q_{\rm abs}$ at $100~\mu$m is satisfied.
In Figure \ref{fig:clr_reach}, we show the colour-colour
relation for this absorption efficiency. The resulting
relation is similar to that in Figure \ref{fig:clr_beta}
with $\beta =1$. Thus, the 60 $\mu$m--100 $\mu$m vs.\
140 $\mu$m--100 $\mu$m colour-colour relation is
determined by the spectral index around
$\lambda\sim 100~\mu$m--140 $\mu$m and is not
sensitive to the details of the absorption
coefficient at the other wavelengths.

\subsection{Silicate-graphite mixture}\label{subsec:mixing}

In reality, both silicate and graphite contribute
to the FIR SED at the same time. Thus, it is worth
examining the FIR colour-colour relation predicted by
mixtures of those two species. As stated in
section \ref{subsec:size}, there is a large uncertainty
in the mixing ratio among models. Thus, we examine
a variety of mixing ratios. In Figure \ref{fig:clr_beta}c,
we show the
FIR colour-colour relation calculated by
a graphite-silicate mixture. First, the mass ratio
of those two species is assumed to be
$\rho_{\rm sil}/\rho_{\rm gra}=1.11$
according to \citet{takagi03}. With this ratio,
graphite dominates the FIR SED and the
FIR colour-colour relation becomes almost the
same as that of graphite (see Figure
\ref{fig:clr_beta}b). A similar
graphite-dominated FIR SED can be seen in
\citet{dwek97}. As the fraction of silicate
becomes larger, the FIR colour-colour relation
approaches that shown in Figure \ref{fig:clr_beta}a.
In order to show this, we also examine a
silicate-dominated composition as proposed for
the LMC. We adopt a mass ratio of
$\rho_{\rm sil}/\rho_{\rm gra}=7.25$.
The result is shown
in Figure \ref{fig:clr_beta}d, which becomes
relatively similar to the silicate colours
shown in Figure \ref{fig:clr_beta}a.
However, it is evident that the colours
of the mixed population are between
those of single species.

\section{Discussion}\label{sec:discussion}

\subsection{Grain optical properties in FIR}
\label{subsec:fir_opt}

We have found in the previous section that the FIR
colours are sensitive to the FIR absorption coefficient
of dust grains and are robust to the absorption
coefficient in wavelengths other than the FIR.
Then we have suggested that the FIR emissivity index of
$\simeq 1$ fits the main correlation. An emissivity
index of $\beta\sim 1$ can also explain
the submillimetre part of the Galactic FIR SED without
introducing a very cold component
\citep{reach95}.\footnote{\citet{lagache98} argue
that the cold component may be due to the presence of
an isotropic (probably cosmic) background. However,
since we only select pixels whose 60 $\mu$m intensity
is larger than 3 MJy sr$^{-1}$, the contamination of
such a background is negligible.} Indeed it would be
possible to explain the FIR colour by mixing two
components with different dust temperatures, but
the robustness of the main correlation against the
Galactic latitudes would strongly require that the
mixing ratio between those two components be uniform
over the entire Galaxy. More natural interpretation
is obviously that the main correlation reflects
the FIR absorption coefficient of dust grains itself.
Thus, it is important to investigate what kind of the
FIR absorption coefficients reproduces the observational
main correlation.

We note that we cannot strongly constrain the
spectral index at wavelength out of the range
treated in this paper and that we do not reject the
possibility that the spectral index varies as a
function of wavelength as we
examined in equation (\ref{eq:reach}). In most of
the previous works on the Galactic FIR spectra,
within the uncertainty, the FIR
SED at $60~\mu{\rm m}\la\lambda\la 200~\mu{\rm m}$
was fitted quite well both
by $\beta =2$ and by $\beta =1$ with different
dust temperatures.
However, since the analysis adopted in this paper
(H06) only uses the pixels with negligible
uncertainty ($<10\%$), we are able to constrain the
FIR dust emissivity index more strongly than
previous works, based on the strong correlation
called main correlation.

Also for extragalactic objects, the FIR spectral index is
a matter of debate. The FIR--submillimetre (sub-mm) SEDs
of dust-rich galaxies such as spiral galaxies usually
show an excess in the sub-mm regime relative to a single
component fit to the FIR regime. This is usually
interpreted as existence of very
cold dust contributing to the sub-mm emission
\citep[e.g.][]{stevens05}. However, it is also important
to stress that the discussion on the amount of very cold
dust strongly depends on the assumed FIR--sub-mm
spectral index. The excess of the sub-mm emission could
be explained by a small spectral index ($\beta$).
\citet{kiuchi04}
show that the FIR--sub-mm SEDs of some galaxies can be
fitted with $\beta\simeq 1$--1.5, similar to our derived
values.

As shown by H06, the main correlation is also consistent
with the FIR colours of nearby galaxies. This implies that
investigating the origin of the main correlation is of
a general significance to the interpretation of the FIR
colours in extragalactic objects.

\subsection{Origin of the sub-correlation}
\label{subsec:mixture}

H06 propose that the sub-correlation shown by the
dotted line in Figure \ref{fig:hibi_fig2} can be explained
by mixing two colours in the main correlation. Such
a mixture of FIR colours in a line of sight is called
overlap effect in H06. 
H06 also show observationally that the sub-correlation
tends to be associated with lines of sight with a high
ISRF intensity indicated by a strong
radio continuum intensity. Thus, the sub-correlation
may be reproduced by a contamination of a FIR
colour with a high radiation field intensity.

In order to investigate the overlap effect in our
framework, we assume a two-component model, in which
two FIR colours calculated with two different radiation
field intensities ($\chi$) are mixed.
Such a superposition of FIR colours is treated
by \cite{onaka07} in a more sophisticated manner,
but our simple approach has an advantage that the
physical quantities are easy to interpret.
We denote the FIR intensity predicted under an ISRF
intensity of $\chi$ as $I_\nu (\lambda ;\,\chi)$.
Then, we examine a two-component model in which
the FIR intensity is composed of a mixture of two ISRF
$\chi_1$ and $\chi_2$ $(\chi_1<\chi_2)$,
$I_\nu (\lambda ;\,\{\chi_1,\,\chi_2\}, f_2)$:
\begin{eqnarray}
I_\nu (\lambda ;\,\{\chi_1,\,\chi_2\}, f_2)
=(1-f_2)I_\nu (\lambda;\,\chi_1)+f_2I_\nu (\lambda;\,
\chi_2)\, ,\label{eq:twocomp}
\end{eqnarray}
where $f_2$ is the contributing fraction of the component
with $\chi=\chi_2$ in the line of sight. Since the
FIR emission is optically thin, the contribution from
each component is proportional to the dust optical depth.
Thus, if the total
dust optical depth at wavelength $\lambda$ is $\tau$,
$f_2\tau$ is occupied with a component with
$\chi =\chi_2$.

\begin{figure}
\begin{center}
\includegraphics[width=8cm]{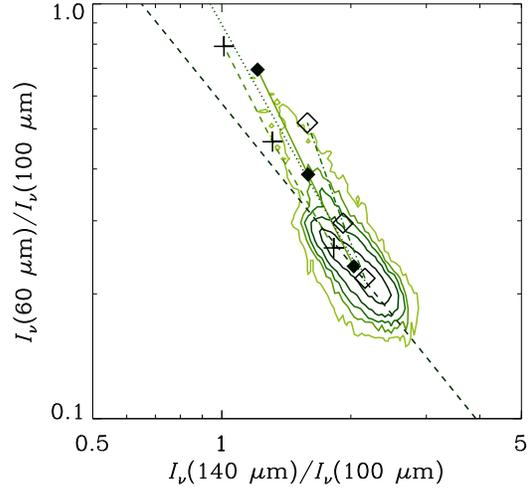}
\end{center}
\caption{Same as Figure \ref{fig:clr_std} but for
the two-component models with $\chi_1=1$ and
$\chi_2=3$, 10, and 30 (lower to upper points).
The filling factors of the second component ($f_2$) are
assumed to be 0.3, 0.1, and 0.03 for the crosses,
filled squares, and open squares, respectively.
\label{fig:highISRF}}
\end{figure}

We adopt the FIR absorption efficiency expressed by
equation (\ref{eq:beta}) with $\beta =1$, since it
explains the main correlation
(section \ref{subsec:emissivity};
Figure \ref{fig:clr_beta}b). We adopt the graphite
species for the absorption efficiency at
$\lambda <100~\mu$m. If we adopt the silicate
absorption efficiency at $\lambda <100~\mu$m, we obtain
similar results, and the following discussions are valid.
We fix $\chi_1=1$ as a general ISRF and examine various
$\chi_2 (>1)$ and $f_2$ as a contaminating high
radiation field. In Figure \ref{fig:highISRF},
we show the results. We find that the main correlation
shifts rightward (i.e.\ toward the sub-correlation)
on the colour-colour diagram if we
mix two colours with different $\chi$. From
Figure \ref{fig:highISRF}, if about 10\% of the total
column of
a line of sight is filled with a high-$\chi$ region,
the sub-correlation is reproduced.
In the following, we discuss the filling factor
assuming that the high radiation field originates from
young massive stars.

We consider typical star-forming regions containing OB
stars as regions with a high radiation field.
\cite{hirashita01} show that a typical Galactic
H {\sc ii} region emit ionizing photons at a rate of
$3.0\times 10^{49}$ s$^{-1}$. If we assume that the
mean energy of the ionizing photons is 15 eV
\citep{inoue00}, the luminosity of the ionizing photons
from the star-forming region becomes
$1.9\times 10^5~L_\odot$. Assuming that the
luminosity of nonionizing photons is
1.5 times as large as that of the ionizing UV photons
\citep{inoue00}, we obtain the nonionizing UV
luminosity of the star-forming region, $\ell_{\rm UV}$,
as $2.8\times 10^5~L_\odot$. We assume that the nonionizing
photons contribute to the dust heating
\citep{buat96,inoue00} and treat the whole stellar cluster
as a point source. Then, the UV radiation flux
($f_{\rm UV}$)
at a distance of $r$ from the centre of the star-forming
region can be estimated as
\begin{eqnarray}
f_{\rm UV} & = &
\frac{\ell_{\rm UV}\,e^{-\tau_{\rm UV}}}{4\pi r^2}
\nonumber \\
& = &
9.0\times 10^{-2}\left(\frac{r}{10~{\rm pc}}\right)^{-2}
e^{-\tau_{\rm UV}}~{\rm erg~s^{-1}~cm^{-2}}\, ,
\label{eq:f_UV}
\end{eqnarray}
where $\tau_{\rm UV}$ is the optical depth of dust grains
in the UV regime. On the other hand, integrating
equation (\ref{eq:isrf}) over the UV regime
($\lambda =912$--4000 \AA), we obtain the UV flux as
$f_{\rm UV}=2.89\times 10^{-3}\chi$ erg s$^{-1}$ cm$^{-2}$
(we used
$7.115\times 10^{-4}\lambda_{\mu{\rm m}}^{-0.6678}$
for $\lambda_{\mu{\rm m}}>0.246$). Equating this with
equation (\ref{eq:f_UV}), we obtain
\begin{eqnarray}
\chi =31\left(\frac{r}{10~{\rm pc}}\right)^{-2}
e^{-\tau_{\rm UV}}\, .
\end{eqnarray}

The UV optical depth ($\tau_{\rm UV}$) is related to the
UV extinction in units of magnitude ($A_{\rm UV}$)
as $\tau_{\rm UV}=A_{\rm UV}/1.086$. If we take
$\lambda =2000$ \AA\ as a representative value
of the UV wavelengths \citep*{hirashita03}, we obtain
$A_{\rm UV}=E_{B-V}(5.52+R_V)=8.60E_{B-V}$ 
for the Galactic extinction curve \citep{pei92}, where
$E_{B-V}$ is the colour excess defined between the $B$ and
$V$ bands ($E_{B-V}\equiv A_B-A_V$),
and $R_V\equiv A_V/E_{B-V}$ ($A_\lambda$ is the
extinction at a wavelength of $\lambda$).
The ratio between the hydrogen column density
(both in atomic and molecular form) and $E_{B-V}$ is
estimated as
$N_{\rm H}/E_{B-V}=5.8\times 10^{21}$ cm$^{-2}$
mag$^{-1}$ in the Galactic environment \citep{bohlin78}.
Combining the
above expressions, we finally obtain the relation
between $\tau_{\rm UV}$ and $N_{\rm H}$ as
$\tau_{\rm UV}=N_{\rm H}/(7.32\times 10^{20}~
{\rm cm}^{-2})$.

\begin{figure}
\begin{center}
\includegraphics[width=8cm]{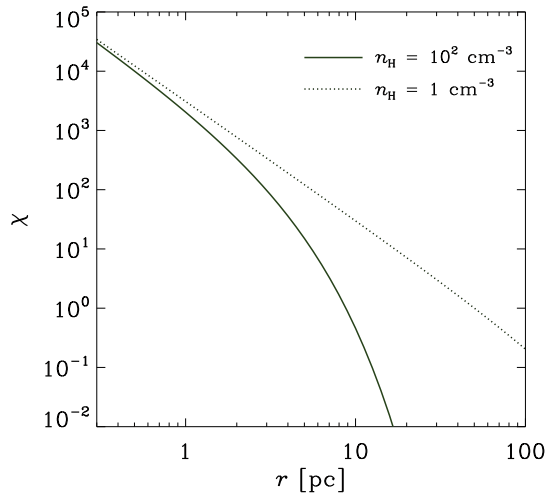}
\end{center}
\caption{ISRF normalized to the solar neighborhood value,
$\chi$, as a function of the distance from the center of
the star-forming region, $r$. The solid and dotted lines
correspond to the results with $n_{\rm H}=10^2$ cm$^{-3}$
and 1 cm$^{-3}$, respectively.\label{fig:isrf_r}}
\end{figure}

\citet{hirashita01} assume that the number density of
hydrogen nuclei in a star-forming region is typically
$n_{\rm H}=10^2$ cm$^{-3}$. Adopting the same number
density, we estimate that $N_{\rm H}=n_{\rm H}r$ (with
$n_{\rm H}=10^2$ cm$^{-3}$) around a star-forming
region. In Figure \ref{fig:isrf_r}, we show the
ISRF as a function of $r$. The sharp decline at
$r\ga 5$ pc is due to the dust extinction.
Although \citet{hirashita01} focused on H {\sc ii}
regions with typical stellar ages less than
$\sim 10^7$ yr, stars with lifetimes of
$\sim 10^8$ yr (B-type stars) also emit UV radiation
\citep[e.g.][]{hirashita01}. Such UV emitting stars
with long lifetimes, after dispersal of dense
regions after $\sim 10^7$ yr,
may be found in the field where the ISM density
is roughly $\sim 1$ cm$^{-1}$. Thus, we also investigate
the radiation field with $n_{\rm H}=1$ cm$^{-3}$. In
this case, the radiation
field declines as $r^{-2}$ and the dust extinction
is negligible up to $r\sim 100$ pc as shown in
Figure \ref{fig:isrf_r}.

H06 have shown that the FIR colour-colour relation follows
the sub-correlation in $\sim 5$--10\% of the lines of sight
in the Galactic plane. The probability (denoted as
$P(r)$) that an arbitrary
line of sight along the Galactic plane passes through a
region with a radius of $r$ can be estimated by
$P(r)\simeq \pi r^2sN_{\rm SF}/(\pi R_{\rm d}^22H_{\rm d})$,
where $s$ is the path length, $H_{\rm d}$ is the scale
height of the Galactic disk, and $N_{\rm SF}$ is the number
of the
regions in the entire Galaxy.
Taking $s\sim R_{\rm d}$, we obtain
\begin{eqnarray}
P(r) & \sim & 0.13\left(\frac{r}{5~{\rm pc}}\right)^{2}
\left(\frac{N_{\rm SF}}{10^4}\right)\left(
\frac{R_{\rm d}}{10~{\rm kpc}}\right)^{-1}\nonumber
\\ & \times &
\left(\frac{H_{\rm d}}{100~{\rm pc}}\right)^{-1}\, .
\label{eq:prob}
\end{eqnarray}
This probability is possibly dependent on the line of
sight through the density structure in the Galactic plane
such as spiral arms, vertical stratification, etc.
However, the above probability is valid if we take a
large number of lines of sight in the Galactic plane
as H06 have done. The estimate of $N_{\rm SF}$ in terms of
the Galactic SFR (or OB star luminosity) is presented in
Appendix \ref{app:sfr}, and the typical SFR (or OB star
luminosity) of the Galaxy gives
$N_{\rm SF}\sim 8\times 10^3$.

The above probability estimated in equation (\ref{eq:prob})
is comparable to the fraction of sightlines with
the sub-correlation if $r\sim 5$ pc. If the sub-correlation
is associated with regions near massive stars, the
sub-correlation passes through a region whose typical
distance from young stars is 5 pc. From
Figure \ref{fig:isrf_r}, the radiation field at
$r\sim 5$ pc from massive stars is typically
$\chi\sim 10$--100. Thus, in order to explain the
sub-correlation, it would be necessary to consider a
mixture of two FIR colours: one with $\chi\sim 1$ as
a normal diffuse ISRF and the other with $\chi\sim 10$--100
as a radiation field near massive stars. This is roughly
consistent with the mixture of the two components
examined above and in Figure \ref{fig:highISRF}.
 
\section{Conclusion}\label{sec:sum}

The observational main correlation between the
60 $\mu$m--100 $\mu$m colour and the
140 $\mu$m--100$\mu$m colour found by H06 is not
explained by the previous models. We have examined
changes of quantities concerning the FIR SED. The most
concentrated region in the colour-colour diagram
and the trend along the main correlation cannot be
explained neither by the change of the ISRF spectrum
nor by the the variation of the grain size distribution.
The FIR colour-colour is not very sensitive to the
heat capacity and the detailed wavelength
dependence of the UV absorption coefficient.
 
On the other hand, we find that the FIR colour-colour
relation is sensitive to the FIR absorption coefficient,
in particular to the emissivity index $\beta$ defined in
equation (\ref{eq:beta}). The observational main
correlation is consistent with the theoretical FIR colours
if we assume that $\beta\simeq 1$--1.5 around
100--200 $\mu$m. This is different
from the spectral index often assumed for
astronomical silicate and graphite species
($\beta\simeq 2$).
Indeed some observational and experimental works
suggest that the FIR spectral index is significantly
smaller than 2 and may be nearly 1
\citep[e.g.][]{reach95,agladze96}.
The difference in the FIR emissivity affects the
estimate of dust mass and of dust temperature.
Thus, we should further examine the FIR dust properties
not only of the Galaxy but also of extragalactic
objects by reexamining FIR data. New data taken by
the {\it Spitzer Space Telescope} \citep{werner04}
and the {\it AKARI}
\citep{murakami04,shibai04,matsuhara06},
both of which have a FIR band with
$\lambda >100~\mu$m, are suitable for such a
purpose.

We have also considered the origin of the sub-correlation
by taking into account the overlap effect; that is,
the FIR radiation comes from regions with a variety of
UV radiation field intensity. For simplicity, we assume
a mixture of two components illuminated by different UV
radiation fields: one is a normal
diffuse component with $\chi\sim 1$ and the other
is an intense field near to young massive stars
($\chi\sim 10$). As a result, we have found that the
sub-correlation can be understood by the FIR colour
with high $\chi$ contaminated by that with normal
$\chi (\sim 1)$. This explains the observational
fact that the sub-correlation is associated with the
regions with intense radiation field (H06). Indeed,
we have also shown that the observed fraction of
the data points on the sub-correlation is consistent
with the total luminosity (or number) of OB stars in
the Galaxy.

\section*{Acknowledgments}
We are grateful to S. Bianchi, R. Schneider, I. Sakon,
Y. Okada, T. Ootsubo, M. Kawada, and the anonymous referee
for their helpful comments. HH has been supported by
Grants-in-Aid for Scientific Research of the Ministry of
Education, Culture, Sports,
Science and Technology (Nos.\ 18026002 and 18740097).
We fully utilized the NASA's Astrophysics
Data System Abstract Service (ADS).


\appendix

\section{Material Dependence}\label{app:mat}

\subsection{Optical constants}

Among silicate species, olivine and pyroxene are often
considered as promising candidates of astronomical
grains \citep[e.g.][]{dorschner95}. Since the optical
constants are different among species, we examine the
robustness of the predicted FIR colour-colour relation
against the change of materials. As for olivine, the
optical constants of two compositions are available:
Mg$_{2y}$Fe$_{2-2y}$SiO$_4$ with $y=0.4$ and 0.5 for
$\lambda =0.20$--500 $\mu$m
\citep{dorschner95}. The optical constants
at $\lambda >500~\mu$m do not affect our results
because of small absorption efficiency,
but those at $\lambda <0.20~\mu$m may be important, since
the absorption of UV radiation determines the dust
temperature. For both $y=0.4$ and 0.5, we adopt the
optical constants of olivine derived by \citet{huffman73}
at $\lambda <0.20~\mu$.

We calculate the FIR colour-colour
($I(60~\mu{\rm m})/I(100~\mu{\rm m})$
vs.\ $I(140~\mu{\rm m})/I(100~\mu{\rm m})$) relation
by adopting the above optical constants. The other
quantities adopted are the same as those in the
text. As a result, we find that the FIR colour-colour
relation of the above materials is the same
as that of astronomical silicate within the
error of the colour-colour relation.
Thus, olivine is not successful in reproducing
the main correlation.
If we adopt the optical constants of pyroxene
found in \citet{dorschner95}
(but we use the olivine optical constants in
\citet{huffman73} for
$\lambda <0.2~\mu$m), the main correlation is not
reproduced. Thus, as far as we examined here,
we conclude that silicate does not
explain the observational FIR colour-colour relation.

We have examined other materials
(FeO, Fe$_3$O$_4$, SiO$_2$, and MgO) that
can be included in silicates by using the optical
constants adopted in \citet{hirashita05}.
We also investigate amorphous carbon instead of
graphite. The optical constants measured by
\citet{edo83} are used
\citep[also adopted by][]{hirashita05}.
However, all of those optical constants do not
change the resulting FIR colour-colour significantly
and do not explain the observational main correlation.
The wavelength dependence of the absorption efficiency
$Q_{\rm abs}$ is given in \citet{hirashita05} and
\citet{takeuchi05} for the above species.

\begin{figure}
\begin{center}
\includegraphics[width=8cm]{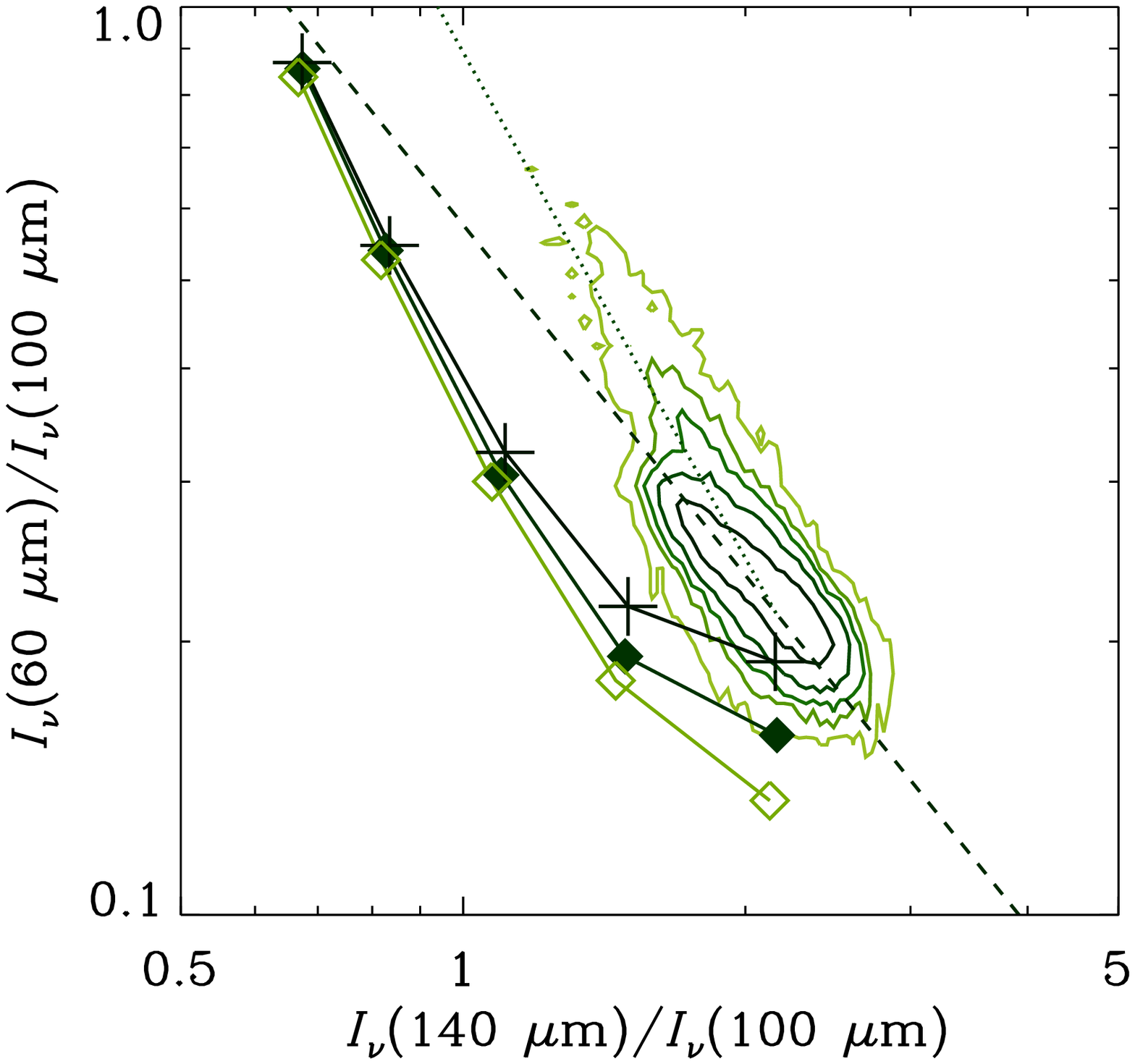}
\end{center}
\caption{FIR colour-colour
($60~\mu{\rm m}-100~\mu{\rm m}$ vs.\
$140~\mu{\rm m}-100~\mu{\rm m}$) relation for various heat
capacities. The other quantities adopted for
calculations are the same as those in
Figure \ref{fig:clr_std}. The filled squares
represent the colour-colour relation with the
heat capacity given in the text for graphite,
and the crosses
and open squares show the results with
heat capacities multiplied by 1/2 and 2,
respectively. The contours and the dashed and dotted lines
are the same as those in Figure \ref{fig:clr_std}.
\label{fig:varc}}
\end{figure}

In our calculation, the optical constants from UV to
FIR are required. The above materials have such data
and are suitable for our work. For completeness, we
should examine all kinds
of possible species, but this is obviously impossible.
However, the above robustness of the predicted FIR
colour-colour relation implies that the detailed wavelength
dependence of the optical constants is not reflected in
the FIR colour-colour relation. We also note that the
wavelength dependence in the FIR absorption efficiency
(emissivity) is roughly proportional to $\lambda^{-2}$
(i.e.\ $\beta\simeq 2$) for all the above species.
As shown in text, the FIR colour-colour relation is
sensitive to the FIR emissivity index. A similar
wavelength dependence of FIR emissivity is the reason why
the above
species produce similar FIR colour-colour relations.

\subsection{Heat capacity}

The difference in the heat capacity may affect the
temperature of the VSGs. The LG temperature is not
affected by the heat capacity since it is determined
by the equilibrium temperature independent of the heat
capacity \citep[e.g.][]{draine84}. Thus, we expect
that the difference in heat capacity may cause a
variation of the 60 $\mu$m intensity.
Here we test heat capacities whose values are 1/2 and
2 times of the value adopted in the text for graphite.
The resulting colour-colour relations are shown in
Figure \ref{fig:varc}. As expected above, the
140 $\mu$m--100 $\mu$m colour, which reflects the
equilibrium temperature, is not affected by the change
of heat capacity. This also supports our assumption
that the contribution from VSGs is negligible at
$\lambda\geq 100~\mu$m. The
60 $\mu$m--100 $\mu$m flux ratio becomes higher
for smaller heat capacity, since the temperature
of a VSG rises more when a photon is absorbed. At the
same time, however, the grains rapidly cool if the
heat capacity is small. Those two effects compensate,
and the 60 $\mu$m--100 $\mu$m colour does not
drastically change. Moreover,
since the contribution
from LGs in radiative equilibrium at the 60 $\mu$m
band becomes larger for higher ISRF, the
difference in the 60 $\mu$m--100 $\mu$m for
$\chi\ga 3$ is negligible. Thus, we conclude that the
change of grain heat capacity does not affect our
results in the text.

\section{The Galactic Star Formation Rate}\label{app:sfr}

In section \ref{subsec:mixture}, we introduce the number
of regions with young massive stars, $N_{\rm SF}$, to
explain the fraction of sightlines associated with the
sub-correlation. The number of such regions reflects
the recent star formation activity of the Galaxy. Thus,
we relate $N_{\rm SF}$ with the Galactic star formation
rate (SFR).

We assume that the Galactic disk has a radius of $R_{\rm d}$
with thickness of $2H_{\rm d}$ (we assume
$R_{\rm d}=10$ kpc and $H_{\rm d}=100$ pc). For simplicity,
we assume that the regions with massive stars distribute
uniformly in the Galactic plane. Then, $N_{\rm SF}$
can be estimated as
$N_{\rm SF}/(\pi R_{\rm d}^2H_{\rm d})$.

The total UV luminosity in the entire galaxy can be
estimated as $N_{\rm SF}\ell_{\rm UV}$, where
$\ell_{\rm UV}$ is the UV luminosity of a region with
massive stars. This can be converted to the SFR as
\citep{inoue00}\footnote{In deriving
equation (\ref{eq:SFR}), we identified
$L_{\rm nonion}=0.6L_{\rm OB}^{\rm bol}$ in
Inoue et al.\ (2000) with
$N_{\rm SF}\ell_{\rm UV}$.}
\begin{eqnarray}
{\rm SFR}=
\frac{N_{\rm SF}\ell_{\rm UV}}{1.8\times 10^9~L_\odot}\,
M_\odot~{\rm yr}^{-1}\, .\label{eq:SFR}
\end{eqnarray}
Since $\ell_{\rm UV}\simeq 2.8\times 10^5~L_\odot$,
we obtain
${\rm SFR}=1.56\times 10^{-4}N_{\rm SF}~M_\odot
~{\rm yr}^{-1}$.

The total luminosity of the Galactic OB stars is
estimated to be $\sim 2.3\times 10^9~L_\odot$
\citep{mathis83}. Equating this with
$N_{\rm SF}\ell_{\rm SF}$, we obtain
$N_{\rm SF}\sim 8.2\times 10^3$ (or
${\rm SFR}=1.3~M_\odot$ yr$^{-1}$). If this number is
used in equation (\ref{eq:prob}), we obtain
the probability that a line of sight is associated
with a high radiation field of $\chi\sim 10$.

\end{document}